\documentclass[pre,twocolumn,amsmath,amssymb]{revtex4}
\usepackage[T1]{fontenc}
\usepackage[latin1]{inputenc}
\usepackage{bm}
\usepackage[dvips]{graphicx}
\usepackage{mathrsfs}
 \usepackage{epsfig}
 \usepackage{psfrag} 
\def\vec{\bm}
\let\SAVEDnabla=\nabla
\def\nabla{\bm \SAVEDnabla}
\def\i{i}

\def\d{\mathrm{d}}
\def\M{\mathscr M}
\def\U{\mathscr U}
\def\L{\mathscr L}
\newcommand{\ord}{{\cal O}}
\newcommand{\bea}{\begin{eqnarray}}
 \newcommand{\ea}{\end{eqnarray}}
\begin{document}
%
\title{Global singularity-free solution of the Iordanskii force problem}
\author{Markus Flaig and Uwe R. Fischer}
\affiliation{Eberhard-Karls-Universit\"at T\"ubingen, Institut f\"ur
  Theoretische Physik\\
Auf der
  Morgenstelle 14, D-72076 T\"ubingen, Germany}
\date{\today}
\begin{abstract}
We present a derivation of the transverse 
force acting on a hydrodynamic vortex in
the presence of an incoming sound wave 
from a global solution of the scattering problem, 
using the method of matched asymptotic expansions. 
The solution presented includes a detailed treatment of the interaction 
of the incident wave with the vortex core, and is free from 
the singularities in the momentum exchange between vortex and sound wave
which have led to contradictory results for the value of 
the transverse force in the literature. 
\end{abstract}
\maketitle
\section{\label{S:intro}Introduction}
The transverse force on a quantized vortex 
due to phonon scattering, first found 
by Iordanskii~\cite{ior66}, has been controversially discussed in 
particular since the work of Thouless, Ao and Niu, which predicted 
that it in fact would vanish \cite{transverse,grisha3,son97, wex,
she98,grishaiord,sto00,son01,tho01,zhang}. 
Based on general properties of 
{\em superfluid} order and the Berry phase associated with the vortex motion,  
they came to the conclusion that 
the total transverse force 
is proportional to the superfluid density alone  
and given solely by the superfluid Magnus force, and not,
as the existence of the Iordanskii force would imply, 
proportional to the  {\em total density}, normal plus superfluid. 
Recent calculations based on the vector potential 
due to the Berry phase of quasiparticles lead to the same conclusion of 
a vanishing Iordanskii force  \cite{zhang}. 

It therefore seems in order to reinvestigate the assumptions 
underlying calculations of the 
Iordanskii force based on classical (but two-fluid) hydrodynamics 
(see, e.g., \cite{son97,sto00}), where 
the transverse force 
stems from the momentum transfer between  
asymmetrically scattered sound waves and the vortex 
\cite{pit59,fet64}. 
The existing treatments 
use various simplifications 
for a treatment of this problem, which constitute sources for 
possible ambiguities and problems in the final result obtained 
for the value of the transverse force. First of all, the authors 
usually consider point vortices. However, 
for small distances to the vortex axis, the corresponding
velocity field cannot be taken literally since then the 
fluid velocity would diverge and the Mach number of
the problem would become infinite. It is not \textit{a priori}
clear that replacing the point vortex by a realistic vortex core would
not affect the oscillatory motion of the vortex core 
due to the incident wave, and thus the momentum
transfer between vortex and sound wave. 
Second, 
using simplified scalar wave equations instead of the full 
linearized hydrodynamic equations,   
one does not take into account the effects due the interaction 
of the incident wave with the vortex core leading to its 
oscillatory motion accurately \cite{son97}. 
Finally, working, as is conventional, 
with an (unphysical) infinitely extended plane wave results in mathematical
difficulties (due to the non-convergence of sums over partial wave
shifts in the scattering cross section \cite{wex}). 
We will show that no such singularities  
appear in the momentum transfer between sound wave and vortex, provided 
one is using a realistically shaped incident wave of a general form.

In the following, we present a study of the scattering problem which avoids
using the simplifications discussed in the above. 
In particular, we derive a {global} solution 
for the coupled hydrodynamic equations 
using the method of matched asymptotic
expansions, matching the solution for the vortical region, with a realistic
core of finite extension, smoothly to the far-field wave region 
solution, assuming a general incoming wave instead of a plane wave. 
As a result, we obtain a regular expression  
for the transverse momentum exchange between sound field and vortex.
When interpreted in terms of two-fluid hydrodynamics, this momentum exchange  
exactly corresponds to the Iordanskii force.

\section{\label{S:basic}Basic hydrodynamic equations}
Our starting point are the Euler and continuity 
equations for an ideal fluid described in terms of density $\rho$, 
velocity $\vec v$ and barotropic pressure functional $P = P (\rho)$: 
\begin{subequations} \label{E:hydro}
\begin{align}
\label{E:continuity}
\frac{\partial \rho}{\partial t} +
\nabla \cdot ( \rho \vec v ) &=
0, \\
\label{E:euler}
\frac{\partial \vec v}{\partial t} +
( \vec v \cdot \nabla ) \vec v &=
-\frac{1}{\rho} \nabla P.
\end{align}
\end{subequations}
In order to consider small perturbations $(\rho_1, \vec v_1)$ on a given
background solution $(\rho_0, \vec v_0)$, which we require to satisfy
the hydrodynamic equations, we introduce the following 
expansions of velocity and density:   
\begin{equation} \label{E:expansion}
\vec v =
\vec v_0 +
\vec v_1 , \qquad 
\rho =
\rho_0 +
\rho_1 
.
\end{equation}
%
Inserting these ans\"atze into~\eqref{E:hydro}, 
and equating terms of first order
on both sides results in the following equations: 
\begin{subequations} \label{E:lin-hydro}
\begin{gather}
\label{E:lin-continuity}
\frac{\d \rho_1}{\d t} + \nabla \cdot ( \rho_0 \vec v_1) +
\rho_1 \nabla \cdot  \vec v_0 = 0, \\
\label{E:lin-euler}
\frac{\d \vec v_1}{\d t} +
(\vec v_1 \cdot \nabla) \vec v_0 =
- \nabla\left(\frac{c^2}{\rho_0} \rho_1\right);
\end{gather}
\end{subequations}
where the sound speed $c$ is as usual defined by 
$c^2 = \partial P/\partial
\rho$ and 
$\d / \d t = (\partial/\partial t + \vec v_0 \cdot
\nabla)$ denotes the convective derivative. 
We now switch from $(\rho_1, \vec v_1)$ to another set of
variables $(\psi, \vec \xi)$ via the following 
decomposition of the perturbations \cite{ber04} 
\begin{equation} \label{E:switch}
\begin{split}
\rho_1 &=
-\frac{\rho_0}{c^2}
\frac{\d \psi}{\d t},\\
\vec v_1 &=
\nabla \psi + \vec \xi.
\end{split}
\end{equation}
The velocity perturbation is decomposed into a 
part with zero vorticity, $\nabla\psi$, where the {\em pressure potential}  
$\psi$ is regular, and a part comprising the vorticity in the sound 
wave, $\vec\xi$. 

Our linearized hydrodynamic equations~\eqref{E:lin-hydro} may
then be rewritten as
\begin{subequations} \label{E:berg}
\begin{gather}
\frac{\d}{\d t} \left(
\frac{1}{c^2}
\frac{\d \psi}{\d t} \right) =
\frac{1}{\rho_0}
\nabla \cdot [ \rho_0 ( \nabla \psi + \vec \xi )],
\label{E:berg-1} \\
\frac{\d \vec \xi}{\d t} +
( \vec \xi \cdot \nabla ) \vec v_0 -
\nabla \psi \times \vec \omega_0 =
0. \label{E:berg-2}
\end{gather}
\end{subequations}
The above coupled pair of equations has first been derived by Bergliaffa
et.\,al.\,\,using Clebsch potentials~\cite{ber04}. Eq.~\eqref{E:berg-1} 
with the vortical part of the fluid velocity perturbations, 
$\vec \xi$, set to zero, was derived in \cite{unr81,pie90}. 
It can be shown ~\cite{ber04}, that in   
the case where the frequency $\varOmega$ of the sound wave is much
greater than the background vorticity $\vec \omega_0 = \nabla \times
\vec v_0$, there exist solutions where $\vec \xi$ is smaller by a 
factor  $\vec \omega_0/\varOmega$ than $\psi$.
This leaves us with 
only one equation for the determination of the scalar quantity $\psi$
when the {\em quasiclassical} condition $\omega_0/\varOmega\ll 1$ 
is fulfilled; see the analysis of the {\em wave region} 
in the section to follow. 

We study an axisymmetric vortex, which may be 
described by the following ansatz for the background solution, 
employing a cylindrical system of coordinates $(r, \phi,
z)$ in what follows: 
\begin{equation} \label{E:background}
\begin{split}
\rho_0 =
\rho_0(r), \qquad 
\vec v_0 = v_0(r) \, \vec e_\phi.
\end{split}
\end{equation}
The radial velocity $v_0$ is linked to the vorticity $\vec \omega_0 =
\omega_0 \, \vec e_z$ (which we consider to be a fixed quantity) 
by the relation 
\begin{equation} \label{E1:vorticity}
v_0(r) =
\frac 1r
\int_0^r
r' \omega_0(r') \, \d r'.
\end{equation} 
We require the vorticity for large $r$ to decay faster
than any power of $r$. This will later on allow us to use an approach
based on the method of matched asymptotic expansions \cite{fed,fls99}. 
For the velocity this assumption implies that, for $r \rightarrow
\infty$, it is given by $v_0 = \gamma/r$ (here and below we ignore
exponentially small terms), with 
$\gamma = \int_0^\infty
r' \omega_0(r') \, \d r'$, which is the circulation divided by $2\pi$. 
While the continuity
equation~\eqref{E:continuity} is automatically satisfied by the
ansatz~\eqref{E:background}, the density $\rho_0$ must be determined
such that the Euler equation~\eqref{E:euler} be fullfilled, which then 
leads to the following equation:
\begin{equation} \label{density}
\rho_0'=
\frac{\rho_0}{c^2}
\frac{v_0^2}{r},
\end{equation} 
where the prime denotes derivation with respect to $r$. 
We immediately see that density variations are of second
order with respect to the Mach number $\M$ which is defined as $\M
= \U/c_\infty$, with $\U$ being some typical velocity in the vortex
and $c_\infty = c(r \rightarrow \infty)$ the sound speed for the fluid at 
rest at infinity. We require the Mach number to be a small quantity ($\M \ll
1$), which allows for an expansion of the  solution 
of the hydrodynamic equations in $\M$ in the section to follow.
\section{Matched asymptotic expansions}
To carry out the matching procedure, we
divide the plane into two regions: An inner,
\textit{vortical} region characterized by $r \simeq \L$, in which the 
vorticity is concentrated, and an outer, \textit{wave} region,
where $r$ is of the order of the typical wavelength 
$\lambda=2\pi \varOmega/c_\infty$ of the incoming wave, 
which we take within the matched asymptotic expansion procedure 
to be $\lambda \simeq \M^{-1} \L$. After having developed suitably 
scaled non-dimensional equations for each region, the solutions in
both region are expanded in the Mach number and matched where the
regions overlap. 

In the vortical region we rescale the spatial variable $\vec r$ according
to $\vec r \rightarrow \L \vec r$ and the velocity $\vec v_0$ according
to $\vec v_0 \rightarrow \U \vec v_0$. The typical time scale of the
problem is given by the inverse of the frequency $\varOmega^{-1}
\simeq \L/\U$.
Consequently, the appropriate scaling for the time
variable $t$ is $t \rightarrow (\L/\U) t$, and the
frequency $\varOmega$ takes the scaled form $(\U/\L) \varOmega$. Thus we
have $|\vec \omega_0| / \varOmega = \ord(1)$, which means, in the light
of the remarks in the previous section,  that we should replace
$\vec \xi$ by $\vec \xi/\L$. With these scalings, and using
Eq.~\eqref{density}, our Eq.~\eqref{E:berg-1} becomes:   
\begin{equation} \label{non}
\nabla \cdot (\nabla \psi + \vec \xi ) = 
\M^2 \left[ \frac{\d^2 \psi}{\d t^2} -
\frac{v_0^2}{r} (\partial_r \psi + \xi_r ) \right] + 
\ord(\M^4),
\end{equation}
while Eq.~\eqref{E:berg-2} stays formally the same.

In the wave region, the appropriately scaled spatial variable is given by $\vec
R = \M \vec r$. Furthermore, we define the velocity potential $\varPsi$
in the wave region as $\varPsi (\vec R) = \psi(\vec r)$. Since in the
wave region the velocity is one order in the Mach number smaller than
in the vortical region, we introduce the wave-region velocity $\vec V_0
= \M^{-1} \vec v_0 = \frac{\gamma}{R} \, \vec e_\phi$. Now, in the
wave region, Eq.~\eqref{non} takes on the following form:  
\begin{equation} \label{NON}
(\nabla^2 - \partial_t^2) \varPsi =
2 \M^2 ( \vec V_0 \cdot \nabla ) \partial_t \varPsi +
\ord(\M^4).
\end{equation}
Here and in what follows, 
we use the convention that the nabla operator acting on a
wave-region quantity corresponds to a derivative with respect to 
$\vec R$. 
The equation for $\vec \xi$ is not needed because we 
restrict ourselves to solutions where in the wave region 
the vortical part of the velocity perturbation, $\vec
\xi$, is exponentially small.

In order to decouple the equations in the vortical region, we split
$\vec \xi$ into an irrotational and a divergence free part according
to \bea 
\vec \xi = \nabla \eta + \vec e_z \times \nabla \zeta, 
\ea 
where $\eta$ and $\zeta$ are uniquely specified by demanding that $\eta,
\zeta \rightarrow 0$ as $r \rightarrow \infty$. Our Eq.~\eqref{non}
then becomes 
\begin{multline} \label{Non}
\nabla^2
( \psi + \eta ) =
\M^2
\left( - i \varOmega + \vec v_0 \cdot \nabla \right)^2
\psi \\
- \M^2 \, 
\frac{v_0^2}{r}
[ \partial_r (\psi + \eta) - r^{-1} \partial_\phi \zeta ] +
\ord(\M^4).
\end{multline}
This Poisson equation may be solved with standard
methods. Using a relation following from 
Eq.~\eqref{E:berg-2} 
\bea
\frac{\partial \vec \xi}{\partial t} 
-\vec v_0 \times (\nabla\times \vec \xi)
+\nabla (\vec v_0 \cdot \vec \xi)
-(\nabla \psi +\vec \xi)\times \vec \omega_0 
= 0 ,\nonumber\\ \label{vecidentity} 
\ea 
and operating with the curl,  
we arrive 
at the following equation: 
\begin{multline} \label{vort}
\left(- i \varOmega +
\frac{v_0}{r}
\frac{\partial}{\partial \phi} \right)
\nabla^2 \zeta 
-\frac{\omega_0'}{r}\frac{\partial \zeta}{\partial \phi} 
\\
= -\omega_0' \frac{\partial}{\partial r} (\psi + \eta) -
\omega_0 \nabla^2 (\psi + \eta) .
\end{multline}
Note that here the fact that 
$\nabla \cdot \vec v_0 = 0$ has been used. The above equation 
allows us to calculate $\zeta$ 
after $\psi + \eta$ has been calculated using~\eqref{Non}. 
%
%

We now formally expand the general solution in the Mach number $\M$
and  then solve the resulting equations to the relevant order in the 
Mach number. In the vortical region the expansion takes on the following form
\begin{equation} \label{expansion}
\begin{split}
\psi &= \psi^{(0)} + \M \psi^{(1)} + \M^2 \psi^{(2)} + \ldots, \\
\eta &= \eta^{(0)} + \M \eta^{(1)} + \M^2 \eta^{(2)} + \ldots, \\
\zeta &= \zeta^{(0)} + \M \zeta^{(1)} + \M^2 \zeta^{(2)} + \ldots;
\end{split}
\end{equation}
while in the wave region we have 
\begin{equation} \label{Expansion}
\varPsi = \varPsi^{(0)} + \M \varPsi^{(1)} + \M^2 \varPsi^{(2)} + \ldots.
\end{equation}
We restrict our analysis to order $\ord(\M^2)$, because at higher
orders the calculations become very involved.
\subsection{The solution to $\ord(\M^0)$}
At this order, the solution in the wave region is determined by the
ordinary wave equation $(\nabla^2 + k^2) \varPsi^{(0)} = 0$,
where the nondimensional wave number $k$ is defined as $k =
\varOmega$. We expand the general solution to zeroth order, corresponding
to our incoming wave, in the following way into partial waves of order 
$\ell$:  
\begin{equation}
\varPsi^{(0)} = \sum_{\ell = -\infty}^\infty A_\ell \, 
J_{|\ell|}(kR) \, e^{i (\ell \phi - \varOmega t)}, \label{Psi(0)}
\end{equation} 
where the $J_{|\ell|}$ are Bessel functions. 
The form of $\varPsi^{(0)}$ in the vortical region may
be determined by expanding the Bessel functions for small $R$, yielding 
\bea
\label{asym}
\varPsi^{(0)} & = & 
A_0 e^{-i\varOmega t} + \M kr \sum_{\ell = \pm 1} \frac{A_\ell}{2}
e^{i (\ell \phi - \varOmega t)} \nonumber\\
& & +\M^2 \frac{k^2 r^2}{4} \left( \sum_{\ell = \pm 2}
\frac{A_\ell}{2} e^{i (\ell\phi - \varOmega t)} -
A_0 e^{-i\varOmega t} 
\right) \nonumber\\
& & + \ord(\M^3).
\ea
%
This will be needed later to carry out the matching.

In the vortical region, Eq.~\eqref{Non} to this order gives the
ordinary Poisson equation $\nabla^2 (\psi^{(0)} + \eta^{(0)}) = 0$,
with the following general solution
\begin{equation}
\psi^{(0)} + \eta^{(0)} =
\sum_{\ell = -\infty}^\infty
a_\ell \,
(kr)^{|\ell|} \,
e^{i (\ell \phi - \varOmega t)}.
\end{equation} 
In the wave region, $r^{|\ell|} = \ord(\M^{-|\ell|})$, ruling out the
existence of terms with $|\ell| > 0$, since these would 
be of \textit{negative} order in the Mach number. Thus, $\psi
^{(0)} + \eta^{(0)} = a_0$.

A decomposition of $\zeta^{(0)}$ into azimuthal modes according to 
$\zeta^{(0)}=\sum_{\ell=-\infty}^\infty\zeta_\ell^{(0)}
(r)\,e^{i(\ell\phi-\varOmega t)}$ reduces 
Eq.~\eqref{vort} to   
\begin{equation} \label{rayleigh}
\left[
\frac{\d^2}{\d r^2} +
\frac 1r \frac{\d}{\d r} -
\frac{\ell^2}{r^2} +
\frac{\ell \omega_0'}{r \varOmega - \ell v_0}
\right] \zeta_\ell^{(0)} (r) =
0.
\end{equation} 
We restrict ourselves to the 
case where there are no critical layers present, 
$\varOmega \neq \ell v_0(r)/r$ for all $\ell, r$~\cite{fls99}. 
In this case, the solutions of this radial Rayleigh equation are 
known to behave for large $r$ as $\zeta_\ell^{(0)} (r) \rightarrow \alpha_\ell\left(
r^{|\ell|} + \beta_\ell r^{-|\ell|} \right)$, immediately giving $\alpha_\ell =
0$ for all $\ell$. 
Furthermore, using identity (\ref{vecidentity}) and 
$v_1^{(0)}=\nabla\psi^{(0)}+ \vec \xi^{(0)}=0$ as well as 
$\nabla\psi^{(0)} =-\nabla \eta^{(0)}$, 
we can derive the equation 
$-i\Omega \eta^{(0)} +\frac{v_0}r \frac{\partial\eta^{(0)}}{\partial\phi }
=$ const., from which it follows 
that $\eta^{(0)}=C e^{i(\Omega r/v_0) \phi}+$ const. Therefore, 
as $\eta^{(0)}$ should be uniquely defined at each point in space,
$C$=0 in the presently considered 
case; 
$\eta^{(0)}$ thus equals a constant, which is zero because   
it must  vanish at large $r$. 

Matching the solution in the vortical region 
with the wave region gives $a_0 = A_0$, thus 
%
\begin{equation} \label{E:zeroth}
\psi^{(0)} = A_0 \, e^{-i \varOmega t}, \quad 
\eta^{(0)} = \zeta^{(0)} = 0.
\end{equation}
To this lowest (zeroth) order in the Mach number, there is thus no
perturbative flow in the vortical region, 
but since $\psi^{(0)} \neq 0$, there are 
density perturbations present.
\subsection{The solution to $\ord(\M)$}
In the wave region, we have once more the ordinary wave equation
$(\nabla^2 + k^2) \varPsi^{(1)} = 0$. We require the solution 
to this equation to be causal; therefore we write  
\begin{equation} \label{hankel}
\varPsi^{(1)} =
\sum_{\ell = -\infty}^\infty 
B_\ell H_\ell^{(1)} (kR)\, 
e^{i (\ell \phi - \varOmega t)}, 
\end{equation}
where the Hankel functions of the first kind $H_\ell^{(1)}$ 
correspond to outgoing cylindrical waves. 
Due to the irregular behavior of the Hankel functions at the
origin, none of the terms in the above sum 
matches to the solution in the vortical region, 
as becomes clear from Eq. (\ref{Asym}) below; therefore, $B_\ell = 0$.  
Thus, $\varPsi^{(1)} = 0$ in the wave region, 
reflecting the simple fact that there
is no sound radiation to linear order in 
the Mach number into the wave region.

In the vortical region, $\psi^{(1)} + \eta^{(1)}$ is once more
determined by the ordinary Poisson equation, giving $\psi^{(1)} +
\eta^{(1)} = \sum_{\ell = -1}^1 b_\ell \, (kr)^{|\ell|} \, e^{i (\ell \phi -
  \varOmega t)}$, where the sum is limited by the fact that higher-order
terms would be of negative order in the Mach number. 
The quantity $\zeta^{(1)}$ ist determined by the following equation:
\begin{multline} \label{E:guess}
\left[ \left(-i\Omega + \frac{v_0}{r} \frac{\partial}{\partial \phi}
\right)\nabla^2 -\frac{\omega_0'}r  
\frac{\partial}{\partial \phi} \right] \zeta^{(1)} 
\\
= -k\omega_0' \sum_{\ell = \pm 1} b_\ell e^{i (\ell \phi - \varOmega t)}.
\end{multline}
Using the ansatz $\zeta_p^{(1)} = f(r) \sum_{\ell = \pm 1} b_\ell 
e^{i (\ell \phi -  \varOmega t)}$ and $\nabla^2 v_0 
= \partial_r (r^{-1} \partial_r (r v_0))= \omega_0'$, 
we immediately see that the equation above is solved once one chooses
$f= -i v_0$. The asymptotics of the solution thus obtained 
for $r \rightarrow \infty$ is given by 
\begin{equation} 
\zeta_\mathrm{p}^{(1)} \rightarrow
-i \frac{\gamma}{r}
\sum_{\ell = \pm 1} b_\ell e^{i (\ell \phi - \varOmega t)}.
\end{equation} 
One gets the full solution to $\zeta^{(1)}$ by adding the general
solution $\zeta_\mathrm{h}^{(1)}$ of the homogeneous Rayleigh equation to
$\zeta_\mathrm{p}^{(1)}$. Since $\zeta_\mathrm{p}^{(1)} \rightarrow 0$ for $r
\rightarrow \infty$, $\zeta_\mathrm{h}^{(1)}\rightarrow 0$ 
for $r\rightarrow\infty$ as well.  
Using the homogeneous version of Eq. (\ref{E:guess}), which can be 
transformed to a Rayleigh equation like Eq. (\ref{rayleigh}); 
the same argument as given after  Eq. (\ref{rayleigh}) then yields 
$\zeta_\mathrm{h}^{(1)}=0$. Since we have
\begin{multline}
\vec e_z \times \nabla \zeta^{(1)} \rightarrow
i \frac{\gamma}{r^2}
\sum_{\ell = \pm 1} b_\ell (\vec e_\phi + i \ell \vec e_r )
e^{i (\ell \phi - \varOmega t)} \\
= \nabla \left[\frac{\gamma}{r}
\sum_{\ell = \pm 1} \ell\,  b_\ell\, e^{i (\ell \phi - \varOmega t)}
\right]
\end{multline}
for $r \rightarrow \infty$, the asymptotics of $\eta^{(1)}$ for $r
\rightarrow \infty$ can be determined to be $\eta^{(1)} \rightarrow
-\frac{\gamma}{r} \sum_{\ell = \pm 1} \ell\, b_\ell\,
e^{i (\ell \phi - \varOmega t)}$, thus giving  
\begin{equation} \label{Asym}
\psi^{(1)}  \rightarrow 
b_0 +
\sum_{\ell = \pm 1} b_\ell \left[ kr + 
\ell \frac{\gamma}{r} \right] e^{i (\ell \phi - \varOmega t)}
\end{equation} 
for $r \rightarrow \infty$. By comparison with the solution in the
wave region, we obtain $b_0 = 0$ and $b_{\pm 1} = A_{\pm 1}/2$. 

The rotational part of $\vec v_1^{(1)}$, 
$\vec e_z\times \nabla \zeta^{(1)}$,  describes the
oscillatory motion of the vortex core induced by the sound wave 
\cite{leb00}, which may be seen as follows. 
Near the origin, we may expand the incoming wave 
into a plane-wave form
$\varPsi^{(0)} \approx \psi_0 e^{i(\vec k\cdot \vec R -\Omega t)}$, 
for which the expansion coefficients are given by 
$A_\ell = \psi_0 e^{- i \ell \phi_k} \,i^{|\ell|}$, with 
$\phi_k$ the angle which the propagation
vector $\vec k = k ( \cos \phi_k,
\sin \phi_k, 0 )$ makes with the $x$ axis. We therefore have    
$\vec e_z \times \nabla \zeta^{(1)}\approx  \vec e_z \times 
(\frac12 \psi_0 \omega_0 \vec e_k) e^{-i\Omega t}$. Then, using
$\vec v_0 \approx \frac12 (\vec \omega_0 \times \vec r) $ near the origin, 
$\vec e_z \times (\frac12 \psi_0 \omega_0 \vec e_k) e^{-i\Omega t}
\approx - \psi_0 (e^{-i\Omega t} \vec e_k\cdot \nabla) \vec v_0 $. 
We see that the term  $\vec e_z \times \nabla \zeta^{(1)}$ 
causes a displacement $\vec r_0(t) \equiv 
\psi_0 e^{-i\Omega t} \vec e_k$ of the vortex line.
It follows for the resulting displacement 
velocity $\frac\partial{\partial t} {\vec r}_0(t) 
= -i \Omega \psi_0 e^{-i\Omega t}\vec e_k 
= \nabla \varPsi^{(0)}|_{\vec r =0}$, so 
that the vortex indeed moves 
with the velocity in the incident sound wave to this order
in the Mach number.
%
\subsection{The solution to $\ord(\M^2)$}
In the vortical region,
$\nabla^2 \left( \psi^{(2)} + \eta^{(2)} \right) =
-k^2 A_0 e^{-i\Omega t}$,
with the following solution:
\begin{equation}
\psi^{(2)} + \eta^{(2)} =
-\frac{A_0 e^{-i\Omega t}}{4} k^2 r^2 +
\sum_{\ell = -2}^2 
c_\ell \, (k r)^{|\ell|} \, e^{i (\ell \phi - \varOmega t)}.
\end{equation} 
%
Comparison
with~\eqref{asym} leads to $c_0 = c_{\pm 1} = 0$ and
$c_{\pm 2} = A_{\pm 2}/8$.
%
In the wave region, $\varPsi^{(2)}$ obeys a forced wave equation:
\begin{equation} \label{E:forced}
( \nabla^2 + k^2 ) \varPsi^{(2)} =
-2 i k \gamma R^{-2} \partial_\phi \varPsi^{(0)}.
\end{equation} 
In the appendix, it is shown that the solution to this equation
may be written as a contour integral.
The result is (with $\cal C$ denoting the 
integration contour shown in the appendix): 
\begin{equation}
\begin{split} \label{solution}
\varPsi_\mathrm{p}^{(2)} 
& =\mathrm{sgn}(\ell)\, i \frac{k \gamma}{2 \pi} A_\ell
\sum_{\ell = -\infty}^\infty 
\varPhi_\ell(R)  e^{i ( \ell \phi - \varOmega t )}
\end{split} 
\end{equation}
where the expansion coefficients are given by 
\begin{equation}
\begin{split}
\varPhi_\ell(R) = 
\int_{\cal C}
\left( \ln t + i \frac{\pi}{2} \right)
\exp \left[i \frac{kR}{2}\left(t - \frac1t \right) \right]
\frac{\d t}{t^{|\ell|+1}}.
\end{split} 
\end{equation}
We can write the general solution to 
Eq.~(\ref{E:forced}) as $\varPsi^{(2)} = \varPsi_\mathrm{p}^{(2)} 
+ \varPsi_\mathrm{h}^{(2)}$, with $\varPsi_\mathrm{h}^{(2)}$ 
being the general solution of the homogeneous equation.
For large $R$, $\varPsi_\mathrm{p}^{(2)}$ corresponds to an outgoing wave
(cf. the appendix). Therefore, $\varPsi_\mathrm{h}^{(2)}$ should also 
contain only the outgoing wave:  
\begin{equation}
\varPsi_\mathrm{h}^{(2)} =
\sum_{\ell = -\infty}^\infty  
C_\ell \, H_\ell^{(1)} (kR) 
e^{i (\ell \phi - \varOmega t)}.
\end{equation}
%
The remaining coefficients are determined by matching to the solution in
the vortical region, giving $C_\ell= 0$ for $\ell\neq \pm 1$ and 
\begin{equation}
C_{\pm 1} = 
\pm i \frac{\pi}{4}
k \gamma A_{\pm 1}.
\end{equation}
%
The solution to quadratic order in the Mach number, 
$\varPsi^{(2)}$, is thus made up of two parts. The first
part, $\varPsi_\mathrm{p}^{(2)}$, may be interpreted to represent 
the interaction
of the incident wave with the long-range velocity field of the vortex,
while the second, $\varPsi_\mathrm{h}^{(2)}$ stands for the interaction of the
wave with the vortex core \cite{fls99}. 
\section{Force on the vortex from phonon scattering}
In this section we will return to the original (dimensional)
variables of section~\ref{S:basic}. Far from the vortex, our solution
takes on the following asymptotical form: 
$\psi = \sum_\ell \psi_\ell (r) \, e^{i (\ell \phi - \varOmega t)}$ where, 
using the asymptotic form of the 
expansion coefficients $\varPhi_\ell(R)$ in 
the far-field solution \eqref{solution},  
derived in the appendix [Eq. (\ref{E:asymptotics})], 
we have 
\begin{equation} \label{form}
\psi_\ell = 
\frac{e^{-i \frac{\pi}{2} |\ell| -i \frac{\pi}{4}}}{\sqrt{2\pi kr}}
A_\ell \left\{
(1 + 2i\delta_\ell ) e^{i kr} +
e^{-i kr + i \pi |\ell| + i \frac{\pi}{2}}
\right\}. 
\end{equation} 
Here, the phase shifts $\delta_\ell$ are given by 
\begin{equation} \label{shifts}
\delta_\ell =
\begin{cases}
\mp \frac{\pi}{4} \frac{k \gamma}{c_\infty} & ; \ell = \pm 1 , \\ 
-\mathrm{sgn}(\ell) 
\frac{\pi}{2} \frac{k \gamma}{c_\infty} & ; \ell \neq \pm 1,   
\end{cases}
\end{equation} 
which are idential to those obtained by Sonin \cite{son97}. 
%
To evaluate the momentum transfer between sound wave and vortex, 
we consider the time-averaged momentum-flux tensor, which reads 
\cite{son97,sto00,stonePRE}:   
%
%
%
%
%
\begin{equation} \label{E:flux}
\langle \varPi_{ij} \rangle =
\left( \frac{c_\infty^2}{\rho_\infty}
\frac{\langle \rho_1^2 \rangle}{2} -
\rho_\infty \frac{\langle v_1^2 \rangle}{2}
\right)
\delta_{ij} +
\rho_\infty \langle v_{1i} v_{1j} \rangle, 
\end{equation}
where $\rho_\infty$ is the constant density far from the vortex. 
We have to insert into this expression 
the time-averaged expressions for the density
and the velocity expressed in terms of $\psi$ and $\vec \xi$ according
to~\eqref{E:switch}. Since we are looking at the region far from
the vortex, we may use the  simplified expressions:
\begin{align}
\rho_1&=
-\frac{\rho_\infty}{c_\infty^2}
{\rm Re} \, [ \partial_t \psi ] =
\frac{i\rho_\infty k}{2 c_\infty} (\psi^* -\psi ), \\
v_{1i} &= 
{\rm Re} \, [ \partial_i \psi ] =
\frac 12 ( \partial_i \psi^* + \partial_i \psi ) ,
\end{align}
which, when inserted into the time-averaged momentum-flux
tensor~\eqref{E:flux}, yield
\begin{align} \label{tensor}
\langle \varPi_{ij} \rangle &=
\frac{\rho_\infty}{4}
[\partial_i \psi^* \partial_j \psi +
\partial_i \psi \partial_j \psi^* 
+( k^2 |\psi|^2 -
|\nabla \psi|^2) \delta_{ij}
].
\end{align}
This comprises a general expression for the time-averaged
momentum-flux tensor in the far-region where $\vec v_0 \approx 0$. 

The projection $ F_{\phi_0}$ of the force on the vortex per
unit length $\vec{ F}$ in the direction of the unit vector 
$\vec n_0 = (\cos \phi_0, \sin \phi_0, 0)$ is given by 
\begin{equation}
 F_{\phi_0} = 
\langle \vec{ F} \cdot \vec n_0 \rangle =
-\oint \langle \varPi_{ij} \rangle n_i \, \d a_j.
\end{equation}
%
Inserting~\eqref{tensor} into the above expression,
and choosing a cylindrical surface around the vortex as 
our contour of integration, 
we get%
\begin{equation}
 F_{\phi_0} = 
-\frac{\rho_\infty}{4}
\int_{-\pi}^\pi \d \phi \, r
\cos ( \phi - \phi_0 ) \, 
( \partial_r \psi^\ast \partial_r \psi +
k^2 \psi^* \psi ).
\end{equation}
Making use of (\ref{form}), and performing the angular integration, we find 
\begin{equation} \label{force}
 F_{\phi_0} =
\rho_\infty k
\!\sum_{\ell=-\infty}^\infty \!
{\rm Re} \left[
i^{|\ell| + 1 - |\ell + 1|} e^{i \phi_0} A_\ell^* A_{\ell + 1}
( \delta_\ell - \delta_{\ell + 1}) \right].
\end{equation} 
Employing the obtained phase shifts for sound-vortex scattering, 
Eq.~\eqref{shifts}, 
we arrive at the following final expression for the force:
\begin{equation} \label{E:result}
 F_{\phi_0} =
\frac{\rho_\infty k^2}{c_\infty} \frac \pi 4 \gamma 
\, {\rm Re} \left[
e^{i \phi_0} 
\sum_{\ell = 0}^1
(A_\ell^* A_{\ell+1} - A_{-\ell-1}^* A_{-\ell})
\right].
\end{equation} 
The above equation represents the general result for the force on the vortex
from phonon scattering, containing partial waves up to order $|\ell|=2$, 
valid for an arbitrary incident wave form. 

Note that there are no singularities in \eqref{E:result} present 
due to partial wave summation, because we have 
assumed an arbitrary incoming wave form right from the start. 
Only now, {\em  after} 
obtaining the general result for the force in terms of the 
expansion coefficients $A_\ell$, we use a plane wave 
approximation for the incoming wave near the origin.
%
 Then, with  $A_\ell = \psi_0 e^{- i \ell \phi_k} \,i^{|\ell|}$, 
\begin{equation} \label{magnus}
 F_{\phi_0} =
-\frac{\rho_\infty k^2}{c_\infty} \pi \gamma |\psi_0|^2
\sin (\phi_0 - \phi_k).
\end{equation}
%
Using the following expression for the momentum density 
of a plane sound wave,
$\vec j = |\psi_0|^2 \frac{\rho_\infty k}{2 c_\infty} \vec k$ 
~\cite{sto00}, Eq. \eqref{magnus} may easily be generalized 
to yield the parallel and transverse components of the force,  
\begin{equation} \label{E:transverse}
\vec{ F}_{||} = 0, \quad 
\vec{ F}_\perp =
-\vec \varGamma \times \vec j,
\end{equation} 
where $\vec \varGamma = 2\pi {\gamma}\vec e_z$ and $\vec j$
is the momentum density of the wave at the 
location of the vortex. The coordinate invariant form of~\eqref{E:transverse}
is obviously not restricted to the case where the vortex axis points into the
$z$-direction. For a vortex in a superfluid we have to
replace $\vec j$ by the thermally averaged momentum
density of the phonons $\vec j_\mathrm{ph}$, which in terms of the
quantities of two-fluid hydrodynamics reads $\vec
j_\mathrm{ph} = \rho_n ( \vec v_n - \vec v_s )$, where 
$\vec v_n$ and $\vec v_s$ are normal and superflow 
velocities, respectively. For a vortex
which is at rest with respect to the superfluid condensate ($\vec v_L =
\vec v_s$) this gives
\begin{equation}
\vec{F}_\mathrm{Iordanskii} =
\rho_n [ \vec \varGamma \times ( \vec v_L -\vec v_n)],
\end{equation} 
which is the Iordanskii force. 
We remark that 
the arguments presented in \cite{sto00} suggest that  
the above form of the force remains invariant 
(to the considered order in the Mach number), also 
if the vortex moves with respect to the superflow. 
\section{Conclusion}
We have solved the problem of scattering of sound waves by a vortex
using an expansion into partial waves and the method of matched 
asymptotic expansions, up to the relevant quadratic order in the Mach number. 
To the best of our knowledge, the 
analytical form of the  solution for the scattered wave 
in the form of a Schl\"afli contour integral, Eq. (\ref{solution}), 
has not been given yet in the literature. 
By evaluating the momentum-flux tensor far from the vortex, we derived  
a general expression for the force exerted by the sound wave on the vortex. 
The force depends only on the lowest partial waves ($|\ell| \le 2$) 
and is thus completely determined by the shape of the incident wave 
in a small region around the vortex axis with a diameter of the order of a few wavelengths 
(an analogous result was obtained by Shelankov for 
an incident wave of beam-shaped form~\cite{she98}). 
We demonstrated that the usage of a
``physical'' incident wave of arbitrary shape (and finite extent) 
removes the mathematical difficulties which plagued previous work 
employing the partial wave method, where
the total transverse cross section, obtained after performing a summation over
all partial waves, contained singularities (was not convergent) 
\cite{son97,transverse,wex}. 
The resulting global, singularity-free solution of the scattering problem
we presented 
then leads, in conjunction with conventional two-fluid hydrodynamics, 
to the Iordanskii force on the vortex, 
confirming its existence as a result of the asymmetrical scattering
of sound waves at a vortex. 

Direct experimental confirmation of the Iordanskii 
force has so far been elusive.
A conceivable experimental procedure to verify
the existence of the Iordanskii force in a superfluid 
is to study the influence of sound pulses 
in Bose-Einstein condensates, created by Bragg scattering 
\cite{BECPhonons}, on the vortex. 
Using Bose-Einstein condensates has, inter alia, 
the advantage of being
able to create sound wave pulses in an accurately controlled manner, with 
variable large momenta of the order of 
$c/\gamma$.  We point out that our approach predicts the transverse force
correctly also if the incoming wave is not a plane wave,  
which will in general be the case in the inhomogeneously 
trapped condensates. If the 
vortex is pinned by an optically created external potential to fix it in 
one position, and, shortly after the sound wave has 
been created, the pinning potential is turned off, 
one should in principle 
be able to observe the transverse displacement of 
the vortex due to the sound wave scattered by it. 

\bigskip

\appendix

\section{\label{A:forced}Forced wave equation}

In this appendix, we derive the solution~\eqref{solution} to the
forced wave equation~\eqref{E:forced}. Employing the ansatz \eqref{solution}, 
%
%
Eq.~\eqref{E:forced} reduces to
\begin{equation} \label{E:Forced}
\left[
R^2 \frac{\d^2}{\d R^2} +
R \frac{\d}{\d R} + k^2 R^2 - \ell^2
\right]
\varPhi_\ell =
-4|\ell| \pi i\,  
J_{|\ell|} (kR).\vspace*{-2.5em} 
\end{equation}
\begin{center}
\begin{figure}[b] 
\centerline{\epsfig{file=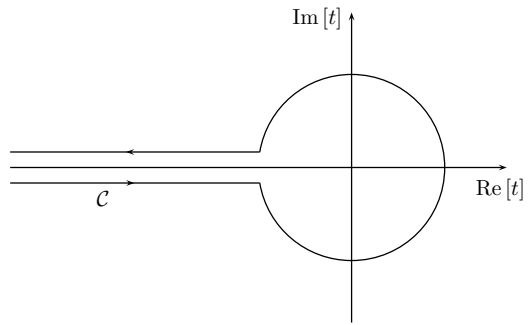,width=0.385\textwidth}}
\caption{\label{F:contour}Schl\"afli contour of integration}
\end{figure}
\end{center}
We write the solution to the above inhomogeneous Bessel 
equation 
as a contour integral.
On the right-hand side, using the Schl\"afli integral representation for
the Bessel functions \cite{Schlaefli}, with the contour $\cal C$ shown in 
Fig.~\ref{F:contour}, we express $J_{|\ell|}(kR)$ as 
\begin{equation} \label{E:schlaefli}
J_{|\ell|} (kR) =
\frac{1}{2 \pi i}
\int_{\cal C} 
\exp \left[\frac{kR}{2}\left(t - \frac1t \right) \right]
\frac{\d t}{t^{|\ell| + 1}}.
\end{equation}
On the left-hand side, inspired by the above Schl\"afli 
integral representation, we try the following ansatz:
\begin{equation} \label{E:ansatz}
\varPhi_\ell =
\int_{\cal C}
g_\ell(t) \, 
\exp \left[\frac{kR}{2}\left(t - \frac1t \right) \right]
\frac{\d t}{t^{|\ell|+1}}.
\end{equation}
After partially integrating the left-hand side and using the well known
recursion relation $J_{\ell-1}(x) + J_{\ell+1}(x) = \frac{2\ell}{x} 
J_\ell(x)$ for
the Bessel functions on the right-hand side, we find that the function
$g_\ell(t)$ is determined by $\frac{dg_\ell}{dt} 
= 
t^{-1}$. If we choose the integration constant such that 
\begin{equation}
g_\ell(t) = 
\ln t + i \frac{\pi}{2} ,
\end{equation}
then the $\varPhi_\ell$ correspond to causal solutions
of~\eqref{E:Forced}. This may easily be verified by applying the
method of steepest descent, yielding the following asymptotics: 
\begin{equation} \label{E:asymptotics}
\varPhi_\ell\rightarrow
-2\pi \sqrt{\frac{\pi}{2k R}}
\exp \left[ i \left(kR - |\ell| \frac \pi 2 - \frac \pi 4
\right) \right]
\end{equation}
for $R \rightarrow \infty$. 
Notice that $\varPsi_\mathrm{p}^{(2)}$ is
regular at the origin, $\varPsi_\mathrm{p}^{(2)} \rightarrow 0$ as $R
\rightarrow 0$, as can easily be confirmed by expanding the integrand for
small $R$ and integrating. 
%

\end{document}